\documentclass[aps,prl,twocolumn,showpacs,floatfix,preprintnumbers]{revtex4}
\usepackage{graphicx}
\usepackage{amssymb}
\usepackage{dcolumn}% Align table columns on decimal point
\usepackage{bm}% bold math
\usepackage{amsmath}
\newcommand{\Mpl}{M_\mathrm{Pl}}
\newcommand{\Ve}{V_\mathrm{eff}}

\newcommand{\bmat}{\beta_\mathrm{m}}
\newcommand{\bgam}{\beta_\gamma}
\newcommand{\rhom}{\rho_\mathrm{m}}
\newcommand{\meff}{m_\mathrm{eff}}
\newcommand{\Pgc}{{\mathcal P_\mathrm{pr}}}%_{\gamma \leftrightarrow \phi}}
\newcommand{\tp}{\tau_\mathrm{pr}}

\newcommand{\Pchamber}{P_\mathrm{chamber}} %pressure in interior of chamber
\newcommand{\Ppump}{P_\mathrm{pump}} %pressure at intake of roughing pump

\newcommand{\fesc}{f_\mathrm{esc}} %zero-bounce escape fraction
\newcommand{\fvol}{f_\mathrm{vol}}%{f_\mathrm{cham}} %V_{chamber}/V_{tot}
\newcommand{\edet}{\epsilon_\mathrm{det}} %total pmt efficiency

\newcommand{\fgam}{F_\gamma}
\newcommand{\faft}{F_\mathrm{aft}}
\newcommand{\ltot}{\ell_\mathrm{tot}}
\newcommand{\Gd}{\Gamma_\mathrm{dec}}
\newcommand{\Agam}{{\vec \Psi}_\gamma}
\newcommand{\Aphi}{{\Psi}_\phi}

\begin{document} 
\preprint{FERMILAB-PUB-08-111-A-CD-E-TD}
\title{A search for chameleon particles using a photon regeneration technique}

\author{A.~S. Chou$^{1}$}
\author{W. Wester$^2$}
\author{A. Baumbaugh$^2$}
\author{H.~R. Gustafson$^3$}
\author{Y. Irizarry-Valle$^2$}
\author{P.~O. Mazur$^2$}
\author{J.~H. Steffen$^2$}
\author{R. Tomlin$^2$}
\author{A. Upadhye$^4$}
\author{A. Weltman$^{5,6}$}
\author{X. Yang$^2$}
\author{J. Yoo$^2$}
\affiliation{
$^1$Center for Cosmology and Particle Physics, New York University, 4 Washington Place, New York, NY 10003\\
$^2$Fermi National Accelerator Laboratory, PO Box 500, Batavia, IL 60510\\
$^3$Department of Physics, University of Michigan, 450 Church St, Ann Arbor, MI 48109\\
$^4$Kavli Institute for Cosmological Physics, University of Chicago, IL 60637\\
$^5$Department of Applied Mathematics and Theoretical Physics, Cambridge CB2 0WA, United Kingdom\\
$^6$Cosmology and Gravity Group, University of Cape Town, Rondebosch, Private Bag, 7700 South Africa
}

\date{\today}
\begin{abstract}
We report the first results from the GammeV search for chameleon particles, which may be created via photon-photon interactions within a strong magnetic field. Chameleons are hypothesized scalar fields that could explain the dark energy problem. We implement a novel technique to create and trap the reflective particles within a jar and to detect them later via their afterglow as they slowly convert back into photons. These measurements provide the first experimental constraints on the couplings of chameleons to photons.

\end{abstract}
\pacs{12.20.Fv, 14.70.Bh, 14.80.Mz, 95.36.+x}
\maketitle

%%%%%%%%%%%%%%%%%%%%%%%%%%%%%%%%%%%%%%%%%%%%%%%%%%%%%%%%%%%%%%%%%%%%%%%%%%%%%%%
\paragraph{Introduction:}

Recent cosmological observations have demonstrated with increasing significance the existence of cosmic acceleration, usually attributed to a negative pressure substance known as ``dark energy''~\cite{frieman08,quint,bp}.  A major effort is under way to discover the properties of dark energy, including its couplings to Standard Model fields.

Perhaps the greatest obstacle to solving this problem lies in the techniques used to probe dark energy. Specifically, our inference of the properties of dark energy have so far come only from observational cosmology data. While the results so far are striking, these techniques are limited in that there are very few, if any, techniques available to separate between different theoretical models of dark energy. This is compounded by the extraordinary difficulty in measuring with enough precision the equation of state parameter w, and its variation in time. 

In this Letter we will illustrate for the first time how dark energy models may be tested in the laboratory.  Unless protected by a symmetry, the dark energy particle should be coupled to all other forms of matter by quantum corrections.  Such couplings can lead to Equivalence Principle violations~\cite{willbook}, fifth forces \cite{fischbach}, variations in Standard Model parameters such as the fine structure constant, and unexpected interactions between known particles. The chameleon mechanism \cite{chamKW1,chamKW2}, by which a matter coupling and a nonlinear self interaction conspire to give a field an environment-dependent effective mass, resolves these issues, while providing a candidate for dark energy.  Crucially, chameleon fields can have small masses on cosmological scales, while acquiring large masses locally in order to evade fifth force searches~\cite{chamKW1,chamKW2,chamcos,GubserKhoury,UpadhyeGubserKhoury,Adelberger2007} while also causing the accelerated expansion observed today. Chameleon dark energy is perhaps most compelling because the very nature of chameleon interactions, if they exist, makes it possible for us to observe them and measure their properties in a diverse array of laboratory tests and space tests of gravity \cite{eotwash, chamKW1, chamKW2}. 

Chameleons can couple strongly to all matter particles with no violations of known physics. Chameleons may also couple to photons via $\phi F^2$ type terms where $F_{\mu\nu}$ is the electromagnetic field strength tensor. Such a coupling allows photon - chameleon oscillations in the presence of an external magnetic field. The chameleon mechanism ensures that a chameleon with large couplings to matter will become massive inside typical laboratory materials.  A chameleon may be trapped inside a ``jar'' if its total energy is less than its effective mass within the material of the walls of the jar.  In this case, the walls reflect the incoming chameleons.  Chameleons produced from photon oscillations in an optically transparent jar can then be confined until they regenerate photons, which emerge as an afterglow once the original photon source is turned off~\cite{gammcham,Ahlers:2007st,Gies:2007su}.  The GammeV experiment in its second incarnation is designed to search for such an afterglow and to measure or constrain the possible coupling of meV mass chameleons to photons.   Probing this low scale in a way complimentary to astrophysics may be the key to understanding the (meV)$^4$ dark energy density.

%%%%%%%%%%%%%%%%%%%%%%%%%%%%%%%%%%%%%%%%%%%%%%%%%%%%%%%%%%%%%%%%%%%%%%%%%%%%%%%
\paragraph{Chameleon phenomenology:}

A chameleon scalar field $\phi$ coupled to matter and photons has an action of the form~\citep{chamcos}
\begin{eqnarray}
S = \int d^4x {\Big (}&-&\frac{1}{2}\partial_\mu\phi\partial^\mu\phi - V(\phi) - \frac{e^{\phi/M_\gamma}}{4}F^{\mu\nu}F_{\mu\nu} 
\nonumber\\
&+& {\mathcal L}_\mathrm{m}(e^{2\phi/M_\mathrm{m}}g_{\mu\nu},\psi^i_\mathrm{m}) {\Big )}
\label{e:action}
\end{eqnarray}
where $g_{\mu\nu}$ is the metric, $V(\phi)$ is the chameleon potential, and ${\mathcal L}_\mathrm{m}$ is the Lagrangian for matter.

For simplicity, we consider a universal coupling to matter  $\bmat = \Mpl / M_\mathrm{m}$, where $\Mpl = 2.4 \times 10^{18}$~GeV is the reduced Planck mass and $M_\mathrm{m}$ is the mass scale associated with the coupling descending from the theory. Theories with large extra dimensions allow matter couplings $\bmat$ much stronger than gravity, while a rough upper bound of $\bmat \lesssim 10^{16}$ is obtained from particle colliders \cite{chamstrong1,chamstrong2}, corresponding to $M_\mathrm{m} > 100$~GeV.  We allow for a different coupling to electromagnetism, $\beta_{\gamma} = \Mpl/M_{\gamma}$, through the electromagnetic field strength tensor $F_{\mu \nu}$.  This term resembles the dilaton-photon coupling $\sim e^{-2\phi} F^2$ in string theory.  

The non-trivial couplings to matter and the electromagnetic field induce an effective potential 
\begin{equation}
\Ve(\phi,\vec x) = V(\phi) + e^{\bmat\phi/\Mpl} \rhom(\vec x) + e^{\beta_\gamma\phi/\Mpl} \rho_\gamma(\vec x),
\end{equation}
where we have defined the effective electromagnetic field density $\rho_\gamma = \frac{1}{2}(|\vec B^2|-|\vec E|^2)$ (for scalars) or $\rho_\gamma = \vec E \cdot \vec B$ (for pseudoscalars) rather than the energy density.  The expectation value $\left< \phi \right>$, the minimum of $\Ve$ and thus the effective mass of the chameleon ($\meff \equiv \sqrt{d^2\Ve/d\phi^2}$ ), depends on the density of both background matter and electromagnetic fields. This dependence is crucial; the afterglow phenomenon requires that the particles have large mass in the walls of the jar (to ensure containment) but that they remain sufficiently light inside the jar to allow coherent, constructive chameleon-photon oscillations over the dimensions of the jar.  For a large range of potentials, the effective mass scales with ambient density as $\meff(\rho) \propto \rho^\alpha$, for $\alpha$ of order unity.  For example, with power law models $V(\phi) \propto \phi^n$ with $n>2$, or chameleon dark energy models ~\cite{chamcos}, $V(\phi) = \Lambda^4 \exp(\Lambda^n/\phi^n)$ with $\Lambda = 2.3$~meV and $n>0$, we find $0 < \alpha < 1$.  Here, $n$ is allowed to be any real number.  As discussed below, our limits on $\beta_\gamma$ will only be valid for models in which the predicted density scaling is strong enough that both the coherence condition and the containment condition can be satisfied.

%%%%%%%%%%%%%%%%%%%%%%%%%%%%%%%%%%%%%%%%%%%%%%%%%%%%%%%%%%%%%%%%%%%%%%%%%%%%%%%
\paragraph{GammeV apparatus:}

\begin{figure}[t]
\begin{center}
\includegraphics[width=2.9in]{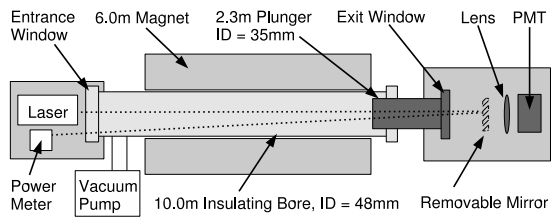}
\caption{The GammeV apparatus. \label{f:chamber}}
\end{center}
\end{figure}

The GammeV apparatus, described in \cite{Chou:2007zzc} and shown in Fig.~\ref{f:chamber}, consists of a long stainless steel cylindrical vacuum chamber inserted into the bore of a $B=5\mbox{ T}$, $L=6\mbox{ m}$ Tevatron dipole magnet.   The entrance and exit of the chamber are sealed with BK7 glass vacuum windows.
A $20$~Hz pulsed  Nd:YAG laser emits $\omega=2.33$~eV photons into the chamber at a rate of $\fgam \sim 10^{19}$~photons/sec.  
The 1 cm$^{-1}$ linewidth of the laser is sufficiently large to span the discrete energy levels of the trapped chameleons.

Interactions with the magnetic field cause each photon to oscillate into a superposition of photon and chameleon states. This superposition can be measured through collisions with the windows; chameleons bounce, while photons pass through.  In order to populate the jar with chameleons, the laser is operated continuously for $\tp\approx 5\mbox{ h}$.  After emerging through the exit window of the chamber, the beam is reflected back through the chamber in order to increase the chameleon production rate and facilitate monitoring of the laser power.  

During the afterglow phase of the experiment, the laser is turned off and a low noise 
photomultiplier tube placed at the exit window is uncovered.  Chameleons interacting with the magnetic field convert back into photons, some of which escape to be detected by the PMT.  Data are taken in two separate runs, with the laser polarization either aligned with or perpendicular to the magnetic field, to search for pseudoscalar as well as scalar chameleons.

Throughout the production and afterglow phases, a pressure $\Pchamber \approx 10^{-7}\mbox{ Torr}$ is maintained inside the vacuum chamber using a turbomolecular pump connected to a roughing pump.  Because the low-mass chameleons are highly relativistic inside the chamber,
the turbo pump simply acts as extra volume ($0.026\mbox{ m}^3$) for the chameleons.  The positive displacement roughing pump is however the weakest ``wall'' of the chamber, and chameleons must be able to reflect on the higher pressure $\Ppump=1.9\times 10^{-3}$~Torr residual gas at the intake of the roughing pump.  The position-dependent $\meff$ acts as a classical potential for the chameleon particle.  A semi-classical tunneling calculation indicates that chameleons will be confined in the chamber over the duration of the data runs ($10^{13}$ reflections) if $(\meff-\omega)>  10^{-6}$~eV at all boundaries of the apparatus.
%Our limits on $\beta_\gamma$ are only valid for models in which the predicted scaling of $\meff$ with density is strong enough that both the above containment condition at the walls and also the coherent oscillation condition  
Also, our experiment is only sensitive to models in which $\meff$ is sufficiently small in the regions away from the walls to allow coherent oscillations: 
%$\meff \ll \mosc \equiv \sqrt{4\pi\omega/L} = 9.8\times 10^{-4}$~eV.
$\meff \ll \sqrt{4\pi\omega/L} = 9.8\times 10^{-4}$~eV.  %away from the walls can be satisfied. % and reflection from the intake of the roughing pump, $\meff > \omega$ at $\Ppump=1.9\times 10^{-3}$~Torr.  
If $\meff$ is dominated by interactions with the residual gas rather than by interactions with the magnetic energy density, then defining $\meff \equiv m_0\cdot (P / \Ppump)^\alpha$, our constraints on $\bgam$ are valid for $\omega < m_0 < \sqrt{4\pi\omega/L} \cdot (\Ppump / \Pchamber)^\alpha$ and hence $\alpha \gtrsim 0.8$ which saturates these inequalities.  %In practice, we can constrain $\alpha > \sim 0.77$ for scalars, and $\alpha > 0.86$ for pseudoscalars with masses that can be probed by our apparatus.
%Otherwise, the range of sensitivity in $\alpha$ is even more restricted. 
Since in our apparatus, $\rhom \approx \rho_\gamma \approx 2\times 10^{-13} \mbox{g/cm}^3$, the experiment is mainly sensitive to models in which $\bmat \gg \bgam$ which in addition predict large density scaling $\alpha$.

%We test our apparatus by generating a diffuse low light level glow discharge within our upstream vacuum region and observing an increased PMT rate.

%%%%%%%%%%%%%%%%%%%%%%%%%%%%%%%%%%%%%%%%%%%%%%%%%%%%%%%%%%%%%%%%%%%%%%%%%%%%%%%
\paragraph{Expected signal:}

In terms of the coupling $\bgam$, and $\meff$ in the chamber, the chameleon production probability \cite{Sikivie:1983ip,Sikivie:1985yu,Raffelt:1987im} per photon is
\begin{equation}
\Pgc = \frac{4\bgam^2B^2\omega^2}{\Mpl^2 \meff^4}
\sin^2\left(\frac{\meff^2 L}{4\omega}\right).
\label{e:Pgc}
\end{equation}
A particle that has just reflected from one of the chamber windows is in a pure chameleon state.  Repeated bounces off of imperfectly aligned windows and the chamber walls cause chameleon momenta to become isotropic.  As a chameleon passes through the magnetic field region, it oscillates between the photon and chameleon states.  In the small mixing angle limit, the photon amplitude $\Agam$ due to this oscillation is given by
\begin{equation}
\left(-\frac{\partial^2}{\partial t^2} - k^2\right) \Agam
= \frac{k\bgam B}{\Mpl} {\hat k}\times(\hat x \times \hat k) \Aphi,
\label{e:Agam}
\end{equation}
where $\Aphi \approx 1$ is the chameleon amplitude, $k\approx \omega$ is the momentum, and $\hat k$ and $\hat x$ are unit vectors in the direction of the particle momentum and the magnetic field, respectively.  The chameleon decay rate corresponding to a particular direction $(\theta,\varphi)$ is $(|\Agam(\theta,\varphi)|^2 + {\mathcal P}_\mathrm{abs}(\theta,\varphi)) / \Delta t(\theta)$ evaluated at the exit window, where $\theta$ is the direction with respect to the cylinder axis, ${\mathcal P}_\mathrm{abs}$ is the total probability of photon absorption in the chamber walls, and $\Delta t(\theta) = \ltot/\cos(\theta)$ is the time required to traverse the $\ltot\approx 12.3\mbox{ m}$ chamber.  We model a bounce from the chamber wall as a partial measurement in which the photon amplitude is attenuated by a factor of $f_\mathrm{ref}^{1/2}$, where $f_\mathrm{ref}$ is the reflectivity.  The mean decay rate $\Gd$ per chameleon is found by averaging over $\theta$ and $\varphi$.  Although the cylinder walls are not polished, a low absorptivity $1-f_\mathrm{ref}=0.1$ is assumed in order to overpredict the coherent build-up of photon amplitude over multiple bounces.  As described below, this overprediction of the decay rate results in a more conservative limit on the coupling constant.  We obtain $\Gd = 9.0\times 10^{-5} (\bgam/10^{12})^2$~Hz.
% for $\bgam=10^{12}$, with $\Gd \propto \bgam^2$.

While the laser is on, new chameleons are produced at the rate of $\fgam \Pgc$ and decay at the rate of $N_\phi \Gd$.  After a time $\tp$ the laser is turned off, and the chamber contains $N_\phi^\mathrm{(max)} = \fgam \Pgc \Gd^{-1} (1-e^{-\Gd \tp})$ chameleon particles.  For our apparatus, this saturates at $3.6\times 10^{12}$ for $\bgam \gtrsim 10^{12}$ and small $\meff$.   The contribution to the afterglow photon rate from non-bouncing chameleon trajectories is
\begin{equation}
\faft(t) = \frac{ \edet \fvol \fesc \fgam \Pgc^2 c}{\ltot \Gd}\left(1-e^{-\Gd\tp}\right) e^{-\Gd t},
\label{E:f_aft}
\end{equation}
 for $t\geq 0$, where $t=0$ is the time at which the laser is turned off.    The detector efficiency $\edet$ contains the $0.92$ optical transport efficiency, as well as the $0.387$ quantum efficiency and $0.7$ collection efficiency of the PMT.  Because chameleons in the turbo pump region do not regenerate photons, we consider only the chameleons in the cylindrical chamber, which represents a volume fraction $\fvol=0.40$ of the total population.  

\begin{figure}[t]
\begin{center}
\includegraphics[width=2in,angle=270]{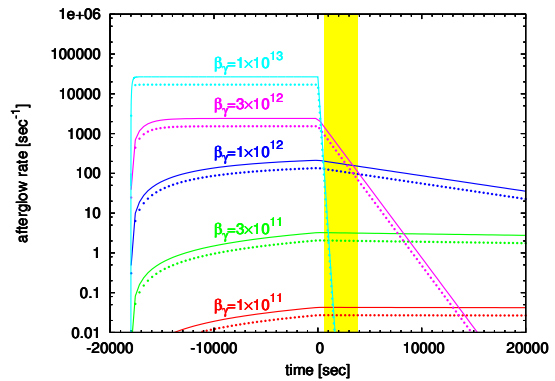}
\caption{Expected afterglow rate for various values of $\beta_\gamma$.  The solid curves are for chameleons with masses of $10^{-4}$~eV while the dotted curves are for $5\times 10^{-4}$~eV chameleons.  Our observation time window for pseudoscalar chameleons is shown shaded in yellow; the corresponding time window for scalar chameleons is shifted to the right by about $700$~sec.\label{F:signal}}
\end{center}
\end{figure}

In order to set conservative, model-independent limits, we consider only the afterglow from the fraction $\fesc=5.3\times 10^{-7}$ of chameleons which travel the entire distance $\ltot$ from entrance to exit windows without colliding with the chamber walls, and are focussed by a 2" lens onto the PMT.  While many chameleons that bounce from the walls may also produce photons which reach the detector (indeed, most of the photons that can reach the detector are on bouncing trajectories), such collisions result in a model-dependent chameleon-photon phase shift \cite{Brax:2007hi} which can affect the coherence of the oscillation on bouncing trajectories.  Figure \ref{F:signal} shows the prediction for the minimum afterglow signal consisting of only the direct light, and attenuated by the fastest possible decay rate $\Gd$ in Eq.~\ref{E:f_aft}.  This afterglow rate is plotted for several values of the photon-chameleon coupling $\beta_\gamma$.  Non-observation of this underpredicted rate sets the most conservative limits.  

%Our goal here is to present results that are, as much as is feasible, independent of the chameleon model and can thus be applied more generally.  We therefore consider only the direct light from non-bouncing trajectories in order to predict the minimum possible afterglow rate for any $\bgam$ and $\meff$.  Furthermore, we use the maximum possible decay rate $\Gd$ in Eq.~\ref{E:f_aft} to allow for the possibility that the afterglow could disappear before we can turn on the detector.  

%However, we also note that in more general models, chameleons could in principle decay to undetectable daughter particles and evade detection.  Furthermore, large chameleon self-interactions can lead to fragmentation and thermalization of chameleons, weakening our constraints.  We leave a more detailed analysis of such effects to a companion paper \cite{gammevcham2}.  

%%%%%%%%%%%%%%%%%%%%%%%%%%%%%%%%%%%%%%%%%%%%%%%%%%%%%%%%%%%%%%%%%%%%%%%%%%%%%%%
\paragraph{Results:}

\begin{table*}[t]
\begin{center}
\caption{\label{F:table} Summary of data for both configurations.}
\begin{tabular}{|l|c|c|c|c|c|c|c|}
\hline
Configuration & Fill Time (s) & $\#$ photons & Vacuum (Torr) & Observation (s) & Offset (s) & Mean Rate (Hz) & excluded (low $\meff$) \\ \hline
Pseudoscalar & 18324 & $2.39\mbox{e23}$ & $2\mbox{e-7}$ & 3602 &  319 & 123 & $6.2\mbox{e11} < \bgam < 1.0\mbox{e13}$ \\
Scalar       & 19128 & $2.60\mbox{e23}$ & $1\mbox{e-7}$ & 3616 & 1006 & 101 & $5.0\mbox{e11} < \bgam < 6.4\mbox{e12}$ \\
\hline
\end{tabular}
\end{center}
\end{table*}

After turning the laser off, we collect afterglow data for one hour on the PMT cathode. Table \ref{F:table} shows relevant data for both of the data runs including: the total integration time during the filling stage, the total number of photons which passed through the chamber, a limit on the vacuum quality (which can affect the chameleon mass and hence the coherence length of the oscillations), the length of the afterglow observation run, the time gap between filling the chamber and observing the afterglow, the mean observed trigger rate, and the limits on $\beta_\gamma$ for coherent oscillations.

In order to minimize the effects of systematic uncertainties due to fluctuations in the dark rate, we compare the expected afterglow signal averaged over the entire observation time to the mean signal observed by the PMT.  The dominant uncertainty in our measurements of the afterglow rate is the systematic uncertainty in the PMT dark rate.  We estimate this quantity, using data from~\cite{Chou:2007zzc}, by averaging the count rate in each of 55 non-overlapping samples approximately one hour in length.  The dark rate, computed by averaging the sample means, is 115 Hz, with a standard deviation of $12.0$~Hz.  No excess is seen in the chameleon data runs over this mean dark rate and all measured rates are well below the $\sim 600$~Hz maximum throughput of our data acquisition system.  The systematic variation in the dark rate is much larger than the statistical uncertainty in the individual sample means.  Thus our $3\sigma$ upper bound on the mean afterglow rate is 36 Hz above the mean of the data rate for each run, after the 115 Hz average dark rate has been subtracted.  

\begin{figure}[t]
\begin{center}
\includegraphics[width=2.in,angle=270]{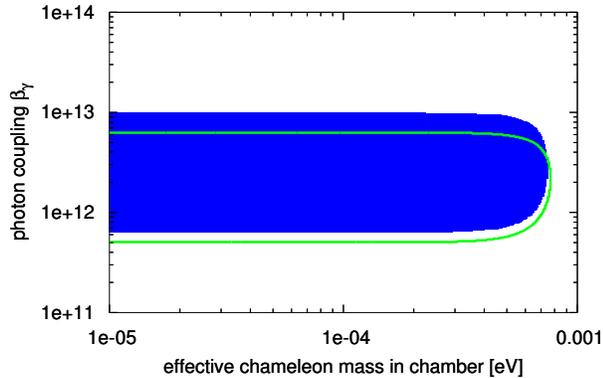}
\caption{Region excluded by GammeV to $3\sigma$ for pseudoscalar particles (solid blue region) and for scalar particles (region between green lines).  %Constraints worsen at $\meff \gtrsim 10^{-3}$~eV as photon-chameleon oscillation becomes incoherent.  %Chameleon interactions with the chamber walls typically prevent $\meff$ from dropping below $\sim 1/R \sim 10^{-5}$~eV. 
%These constraints are valid only for chameleon models in which the mass scales quickly enough with background density that $\meff< 10^{-3}$~eV in the interaction region increases to $\meff>2.33$~eV at the intake of the roughing pump, $P_\mathrm{rough} = 1.9\times 10^{-3}$~Torr, so that chameleons cannot escape the chamber.  Also, for power law chameleon potentials, the chameleon coupling to the walls of the $R$=24mm radius chamber contributes a quantity of order $1/R \sim 10^{-5}$~eV to the chameleon mass inside the chamber~\cite{UpadhyeGubserKhoury,Brax:2007hi}, so that the total effective mass cannot be made lower than this.  %In regions near the walls of the apparatus, the larger effective masses ($> 10^{-3}$~eV) render the oscillations incoherent.  
%The resulting regions of diminished regeneration efficiency are negligibly
%small compared to the total volume of the chamber.
\label{F:results}}
\end{center}
\end{figure}

For each $\meff$ and $\bgam$ we use (\ref{E:f_aft}) to compute the total number of excess photons predicted within the observation time window.  Figure~\ref{F:results} shows the regions excluded by GammeV in the $(\meff,\bgam)$ parameter space for scalar and pseudoscalar chameleon particles.  At $\meff$ near $\sqrt{4\pi\omega/L} = 9.8\times 10^{-4}$~eV, our exclusion region is limited by destructive interference in chameleon production.  
%Although there are peaks in the predicted signal at higher $\meff$, all of them fall below our detection threshold.  
At higher $\meff$, a larger $\bgam$ is needed to produce an equivalent non-bouncing minimum signal rate.  However,
for $\bgam \gtrsim 10^{13}$ our sensitivity diminishes because, as shown in Fig.~\ref{F:signal}, the chameleon decay time $\Gd^{-1}$ in GammeV could be less than the few hundred seconds required to switch on the PMT.  
%As shown in Fig.~\ref{F:signal}, for $\bgam \gtrsim 10^{13}$ the chameleon decay time $\Gd^{-1}$ in GammeV is less than the few hundred seconds required to switch on the PMT, and our sensitivity diminishes.  
Our constraints could be extended to higher $\bgam$ by more quickly switching on our detector, by reducing the magnetic field strength, and/or by making the chamber walls less reflective to reduce $\Gd$.  Finally, at low $\bgam$ we are limited by our uncertainty in the PMT dark rate.  At low $\Gd$, Eq.~\ref{E:f_aft} reduces to a constant rate $\faft \approx \edet \fvol \fesc  \fgam \Pgc^2 c / \ltot$, which, for $\bgam \lesssim 5\times 10^{11}$, is below our detection threshold.  In summary, GammeV has carried out the first search for chameleon afterglow, a unique signature of photon-coupled chameleons.  Figure~\ref{F:results} presents conservative constraints in a model-independent manner, over a restricted range of chameleon models. Improvements to this experimental setup have the potential to open up the chameleon parameter space to testability. Hopefully, this work will inspire others to consider alternative ways to test for dark energy - in high and even low energy experiments. 

\paragraph{Acknowledgements:}
We thank the staff of the Fermilab Magnet Test Facility of the Fermilab Technical Division, and the technical staff of the Fermilab Particle Physics Division, who aided in the design and construction of the apparatus.  We are grateful to S. Gubser and W. Hu for many informative discussions.  JS thanks the Brinson Foundation for their generous support.  This work is supported by the U.S. Department of Energy under contract No. DE-AC02-07CH11359.  AC is supported by NSF-PHY-0401232.

%%%%%%%%%%%%%%%%%%%%%%%%%%%%%%%%%%%%%%%%%%%%%%%%%%%%%%%%%%%%%%%%%%%%%%%%%%%%%%%

\bibliography{gammevcham1}{}

\end{document}